\documentclass{ws-procs975x65}

\newcommand{\be}{\begin{equation}}
\newcommand{\ee}{\end{equation}}
\newcommand{\bea}{\begin{eqnarray}}
\newcommand{\eea}{\end{eqnarray}}
\newcommand{\bml}{\begin{subequations}}
\newcommand{\eml}{\end{subequations}}
\newcommand{\bfig}{\begin{figure}}
\newcommand{\efig}{\end{figure}}

\begin{document}
$~~~~~~~~~~~~~~~~~~~~~~~~~~~~~~~~~~~~~~~~~~~~~~~~~~~~~~~~~~~~~~~~~~~~~~~~~~~~~~~~~~~~$\textcolor{red}{\bf TIFR/TH/15-48}
\title{Cosmic Hysteresis}

\author{Sayantan Choudhury}

\address{Department of Theoretical Physics, Tata Institute of Fundamental Research, \\ 
Mumbai 400005, India.\\ E-mail: sayantan@theory.tifr.es.in, sayanphysicsisi@gmail.com}
\author{Shreya Banerjee}

\address{Department of Astronomy and Astrophysics, Tata Institute of Fundamental Research, \\ 
Mumbai 400005, India.\\ E-mail: shreya.banerjee@tifr.res.in}

\begin{abstract}
{\it Cosmological hysteresis}, has interesting and vivid implications in the scenario of a cyclic bouncy universe. This, purely thermodynamical in nature,
is caused by the asymmetry in the equation of state parameter during expansion and contraction phase of the universe, due to the presence of a single scalar field. When applied to variants of modified 
gravity models this phenomenon leads to the increase in amplitude of the consecutive cycles of the universe, provided we have physical
mechanisms to make the universe bounce and turnaround. This also shows that the conditions which creates
a universe with an ever increasing expansion, depend on the signature of $\oint pdV$ and on model parameters.
\end{abstract}

\keywords{Cosmological hysteresis; Cyclic cosmology; Bouncing cosmology; Cosmology beyond the standard model.}

\bodymatter

\section{Introduction}
Hysteresis is a phenomenon,  occurring naturally in several magnetic and electric
materials in condensed matter physics. In analogy with magnetic and electric hysteresis,
in cosmology we have the phenomenon of cyclic universe (Refs.~ \refcite{eliade,jaki,starobinsky}). But
a universe with identical cycles is unable to solve Big Bang conundrums. However
Tolman in his paper given in Ref.~\refcite{tolman} postulated that the presence
of a viscous fluid would create inequality between the pressures at the time
of expansion and contraction phases which would result in the growth of
both energy and entropy. This unusual approach, though resulted in an inevitable increase
in entropy, helped in solving the horizon and
flatness problem due to the creation of an oscillating universe with increasing
expansion maximum after each cycle. Several other models as in Refs.~ \refcite{Baumann:2014nda,Baumann:2009ds,Lyth:1998xn} were
also proposed to solve the horizon and flatness problem, but none of them could
avoid big bang singularity. However, in Refs.~ \refcite{Kanekar:2001qd,Sahni:2012er}, the
authors created the asymmetry in pressure using the scalar field dynamics, thereby
maintaining the symmetric nature of the equation of motion hence
avoiding entropy production. This led to the production of hysteresis
defined as $\oint pdV$, during each oscillatory cycle. The loop area, hence asymmetry, is largest
in case of inflationary potentials. But the phenomenon of hysteresis is 
independent of the nature of potential. Any potential with proper minimum/minima which
randomizes the phase of the scalar field during
expansion, is capable of causing the phenomenon of hysteresis. In order to avoid singularity
we apply this phenomenon to models where singularity is
replaced by bounce and big crunch replaced by re-collapse or turnaround.

In this paper, we have investigated the phenomenon
of hysteresis in Einstein Gauss-Bonnet (EHGB) brane world gravity
model. For a complete analysis for other models one may refer 
to Ref.~\refcite{Choudhury:2015baa}. A notable feature
of this analysis is that an increase in expansion maximum after each cycle now
depends on the sign of $\oint pdV$ and also on the parameters of
the EHGB model. Thus this analysis helps us to put constraints
on the EHGB model parameters in the bouncing scenario
along with cosmological hysteresis. Though the analysis that
we have performed holds good under certain physically acceptable approximations and limiting cases, we can at least show that if there are any limiting case of EHGB model
which can give rise to the phenomenon of cosmological hysteresis. 

\section{Generating Cosmic Hysteresis}
\label{sq1}
We know, in the spatially flat FLRW cosmological background the scalar field equation of motion is given by:
$\ddot \phi + 3 H \dot \phi + V^{'}(\phi) = 0 \,.$
Here $H$ is the Hubble parameter and the second term $3 H \dot \phi$ acts as friction and opposes the motion of the scalar field, thus producing a  
damping effect during its motion, when the universe expands ($H>0$). By contrast, when the Universe enters a contracting ($H<0$) phase, the term $3 H \dot \phi$ behaves like 
anti-friction and favours the motion of the scalar field, thereby accelerating it. This gives rise to different pressures during contraction and expansion of the Universe, hence to the phenomenon of ``cosmological hysteresis''.
\begin{figure}[ht]
\centering
\includegraphics[width=7cm,height=4cm]{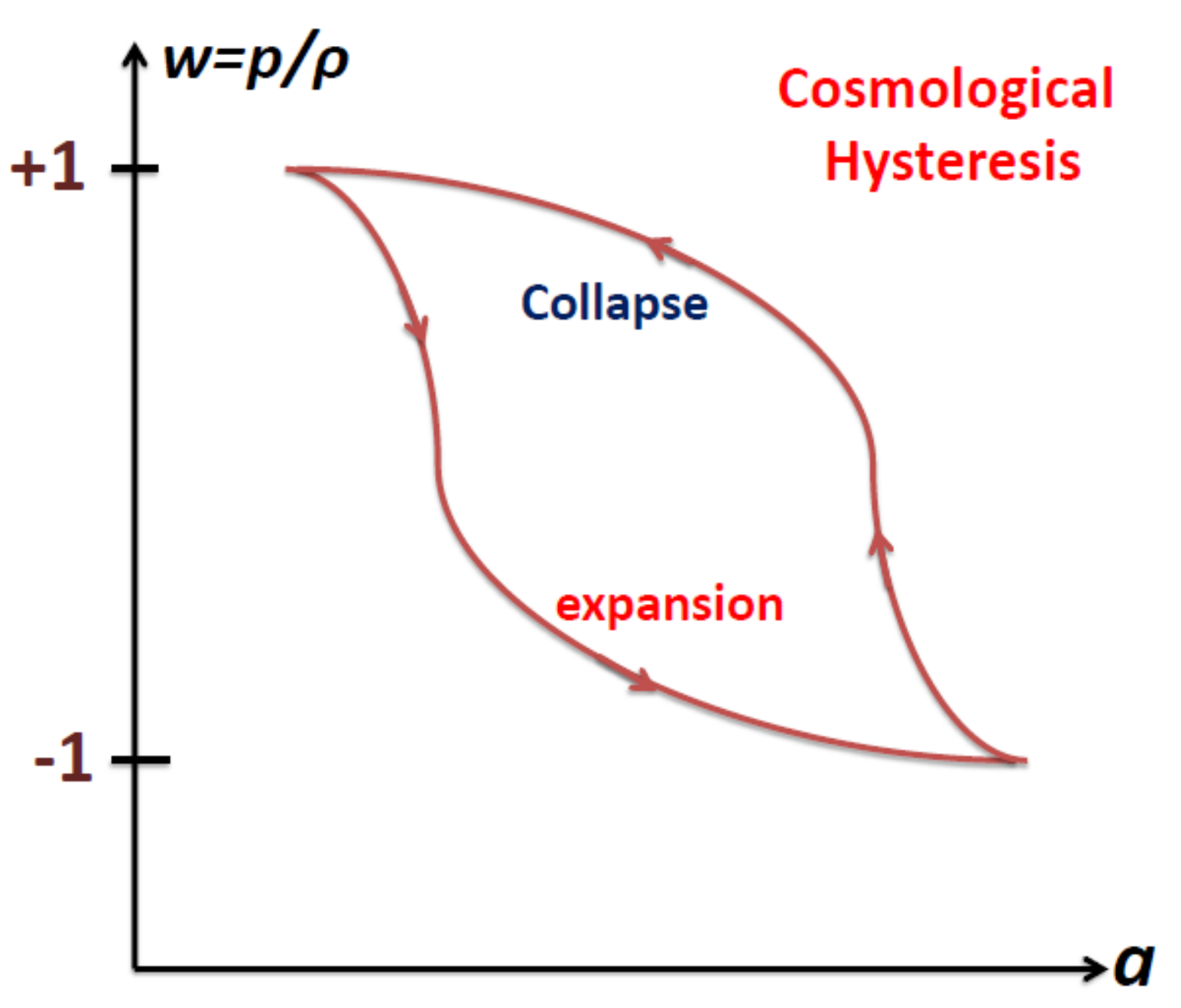}
\caption{\small An idealised illustration of cosmological hysteresis. The loop has been plotted in the $w-a$ plane.  
}
\label{fig:wa}
\end{figure}
It has been first pointed out in Ref.~\refcite{Kanekar:2001qd} that when we plot the equation
of state given by $w = p/\rho$ vs the scale factor $a$ from a specified cosmological model,
we get a hysteresis loop whose area contributes
to the work done by/on the scalar field during expansion and contraction of the Universe.
The general expression for the work done by/on the scalar field during one cycle is given
by $\oint pdV = \int_{cont} pdV + \int_{exp} pdV$.
 The signature of the integral depends on the relative pressure difference  between the contraction and expansion phase i.e $ p_{cont} > p_{exp}$, then the overall signature of the $pdV$ work
 done is negative or, $\oint pdV < 0$. and vice versa. Fig.~\ref{fig:wa} shows the graphical plot between the equation of state
and scale factor of the universe. It illustrates the phenomenon of cosmological hysteresis.

Now following the proposal of  Ref.~\refcite{Kanekar:2001qd,Sahni:2012er},
 we know that in order to get a cyclic universe, the condition for bounce
 and turn around are given by: {\it Bounce}- $H=0$ and $\ddot a> 0$, {\it Turn around}- $H=0$ and $\ddot a< 0$.
 \begin{figure}[ht]
\centering
\includegraphics[width=8cm,height=3cm]{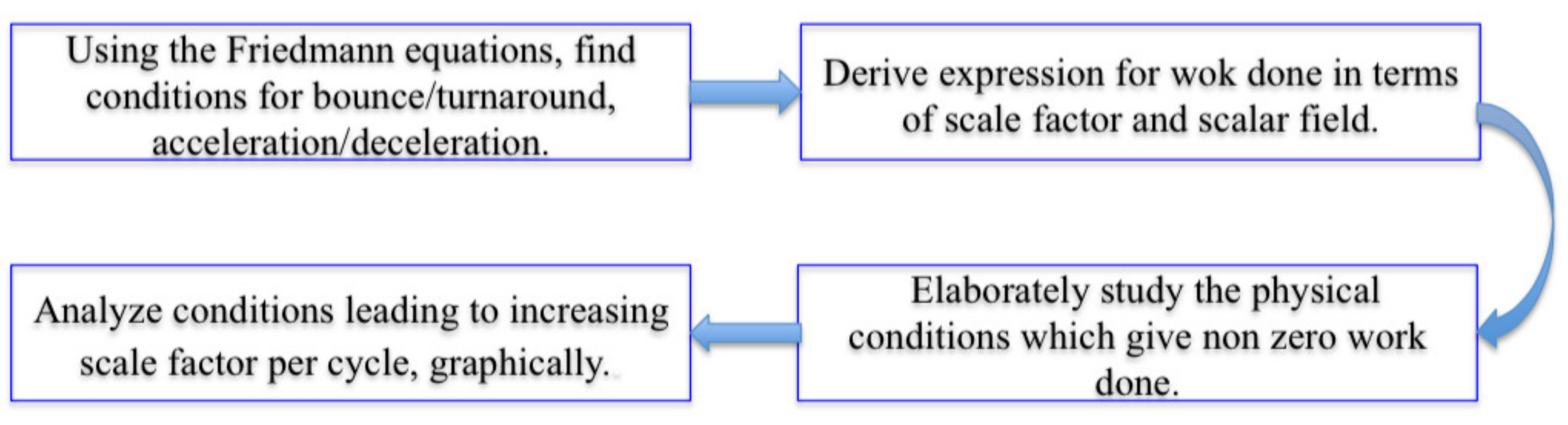}
\caption{\small A schematic representation of the basic steps in our analysis. In this paper, we have discussed the first two steps in some detail.  
}
\label{fig:algorithm}
\end{figure}
In this paper we will discuss the analysis leading to bounce, turnaround and hysteresis for EHGB model. Fig.~\ref{fig:algorithm} shows the basic methodology/steps for our present analysis schematically. In this paper, we have discussed the first two steps in detail for EHGB model. The rest have been discussed in Ref.~\refcite{Choudhury:2015baa}.
\section{Hysteresis from EHGB braneworld}
The modified Friedmann equation in this model is given by (Ref.~\refcite{Maeda:2007cb}):
\begin{equation}
\frac{\kappa^4_5}{36}(\rho+\sigma)^2 = \left(\frac{h(a)}{a^2}+\varepsilon
H^2\right)\left[1+\frac{4\alpha}{3}
\left(\frac{3k-\varepsilon h(a)}{a^2}+2H^2\right)\right]^2\, , \label{GB-F}
\end{equation}
where $\sigma$ is the single brane tension, $\alpha$ is the Gauss-Bonnet coupling,
$\Lambda$ is the 5-D cosmological constant, $\varepsilon = +1,-1$ for space-like or time-like extra dimension respectively. Here the the function $h(a)$ is defined as \be h(a) = \varepsilon k+\frac{a^2}{4\alpha}\left(\varepsilon\mp
\sqrt{1+\frac{\alpha \mu}{a^{4}}+\frac43\alpha\Lambda}\right),\ee where $\mu$ is a constant and $k=0,\pm 1$.

Following the analysis of Ref.~\refcite{Maeda:2007cb}, we can rewrite the above equation in a simplified manner as:
\begin{equation} \label{plus1}
C (\rho + \sigma)^{2} = \left(A \pm H^{2}\right) \left( B + H^{2} \right)^{2} \, ,
\end{equation}
where $\pm$ corresponds to $\varepsilon = +1$ and $\varepsilon = -1$ respectively. Here $A,B$ and $C$ are defined as:
\begin{eqnarray}
A= \frac{k}{a^{2}} + \frac{h(a)}{a^{2}} \, ; B= \frac{3k}{2a^{2}} + \frac{3}{8\alpha} - \frac{\varepsilon h(a)}{2a^{2}} = \frac{3}{8\alpha} + \frac{3k}{2a^{2}}
- \frac{\varepsilon A}{2} \, ; C= \frac{\kappa_5^4}{36} \left( \frac{3}{8 \alpha} \right)^2 > 0 \,.~~~~
\label{ABC}
\end{eqnarray}

Below we have discussed the necessary conditions leading to bounce and turnaround for space-like extra-dimension only. The analysis for time-like extra-dimension is given in Ref.~\refcite{Choudhury:2015baa}.
\begin{itemlist} 
\item \textbf{\underline{Condition for bounce}:}  Bounce occurs when the universe
reaches its minimum radius $a_{min}$ and maximum density $\rho_{b}$, which in this setup is achieved when \be \rho_{b} = \frac{\sqrt{A}B}{\sqrt{C}} - \sigma.\ee
 The expression for change in amplitude of the scale factor at each successive cycle is given by:
\be \delta a_{min} = \frac{\oint pdV}{\left(3\sigma a_{min}^{2} - X'\right)}.\ee Here $X'$ is a new model dependent parameter which is a function of $A,B$ and $C$ (for complete expression, one may refer to Ref.~\refcite{Choudhury:2015baa}). Thus we see that the condition for an increase in the amplitude of the scale factor depends on $A$ through $X'$, which in turn depends on the amplitude and sign of the curvature parameter $k$ and the model parameters like $\mu,\ \alpha$ etc.
 \begin{figure}[ht]
\centering
\includegraphics[width=7cm,height=4.5cm]{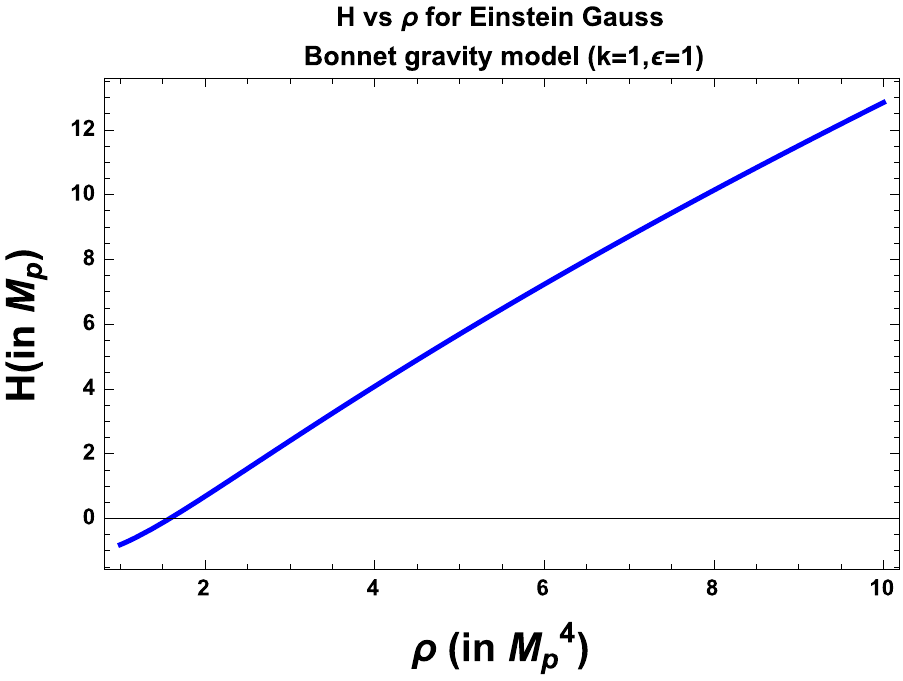}
\caption{\small An illustration of the bouncing condition for a universe with $k=1,\ \varepsilon=1,\ w=1/3,\ C=100,\ A=10M_{p}^{2},\ B=40M_{p}^{2},\ \sigma=-10^{-9}M_{4}^{4},\ C=100,\ \Lambda=6{\rm \times}10^{-3}M_{p}^{2}$.  
}
\label{fig:egb2}
\end{figure}
In Fig.~\ref{fig:egb2}, we have shown an example illustrating the phenomenon of bounce for $k= +1$ in this setup.

\item \textbf{\underline{Condition for acceleration}:} The second Friedmann equation in this context is given by:
 \begin{equation}
 \dot{H}+H^2=\frac{\ddot{a}}{a}  = -\frac{3\sqrt{C}^{2}(\rho + p)(\rho + \sigma)}{Y} + \left(\frac{\dot{a}}{a}\right)^{2} - \frac{Z}{Y}
 \label{gbacceleration}
 \end{equation}
 where $Z$ and $Y$ are functions of $A, B$ and their time derivatives. 
 There explicit expressions have been shown in Ref.~\refcite{Choudhury:2015baa}. The condition for acceleration at bounce is \be p_{b} < \frac{\sqrt{A}B}{\sqrt{C}}\left(-\frac{Z}{3AB^{2}} - 1 + \frac{\sigma \sqrt{C}}{\sqrt{A}B}\right).\ee
 Thus it implies that whether the condition for acceleration violates the energy condition, now depends upon the values of different parameters of the EHGB model present in the above expression.

\item \textbf{\underline{Condition for turnaround}:} Turnaround or re-collapse occurs when the Universe reaches its maximum radius $a_{max}$ and minimum density. Following the same line of treatment as for bounce, the condition for turnaround is \be \rho_{t} = \frac{\sqrt{A}B}{C'} - \sigma.\ee
The expression for change in amplitude of the scale factor \be \delta a_{max} = \frac{\oint pdV}{(3\sigma a_{max}^{2} - X)}.\ee Thus the conclusions for turnaround remains same as that for bounce. 

\item \textbf{\underline{Condition for deceleration}:} The mathematical expression reflecting the necessary condition for deceleration remains same as that
for acceleration and is given by \be p_{t}> \frac{\sqrt{A}B}{\sqrt{C}}\left(-\frac{Z}{3AB^{2}} - 1 + \frac{\sigma \sqrt{C}}{\sqrt{A}B}\right).\ee Just like the case of acceleration within EHGB setup, the condition for deceleration
at turnaround also depends on the EHGB model parameters. 

\item \textbf{\underline{Evaluation of work done in one cycle}:} 

The general mathematical expression for the work done in one complete cycle (i.e contraction$\rightarrow$expansion$\rightarrow$contraction) is
\begin{equation} 
\oint pdV = \int_{a_{max}^{i-1}}^{a_{min}^i-1} 3\left(\frac{\dot{\phi}^{2}}{2} - V(\phi)\right)a^{2}\dot{a}dt + \int_{a_{min}^{i-1}}^{a_{max}^i} 3\left(\frac{\dot{\phi}^{2}}{2} - V(\phi)\right)a^{2}\dot{a}dt
\end{equation} 
where $i$ and $(i-1)$ refer to the two successive cycles i.e. $i$th and $(i-1)$th cycle of expansion and contraction phase of the Universe. Using the Friedmann equations for EHGB model, one can get the corresponding expression for work done in one cycle for this model. The complete expression has been shown in Ref.~\refcite{Choudhury:2015baa}, from which we can conclude that the work done now depends not only on the scale factor, but also on the different parameters of the model like the coupling constant, brane tension etc. within the present setup.
\end{itemlist}

Exact expressions for the evolution of the scale
factor and scalar field with time can be obtained by
solving the Friedmann equations under certain limiting
conditions and valid approximations. It can also be shown
that we get a non zero expression for the total work
done in one cycle under such conditions. For more details see Ref.~\refcite{Choudhury:2015baa}. 

\section{Conclusion}

Through our present analysis we see that the phenomenon of hysteresis for 
EHGB (and other) model(s), is very robust and eternal, which almost naturally, makes us appreciate and acknowledge
its importance in cosmology. Cosmic hysteresis is important due to the
simplicity with which it can be generated. Only a thermodynamic interplay between the pressure
and density, in the presence of a scalar field, succeeds in causing cosmic hysteresis. In the present work and in Ref.~\refcite{Choudhury:2015baa} we study essentially those effective field theoretic models which give rise to both the conditions for bounce and turnaround and at the same
time also satisfy the observed features of the present universe. In future we plan to connect these analyses with CMB, study perturbations, non-Gaussianities and structure formation in this scenario. 

\section*{Acknowledgments}
	SC would like to thank Department of Theoretical Physics, Tata Institute of Fundamental
Research, Mumbai for providing the Visiting (Post-Doctoral) Research Fellowship.  SC takes this opportunity to thank sincerely to Prof. Sandip P. Trivedi and Prof. Shiraz Minwalla
for their constant support and inspiration.  
SC would like to acknowledge our debt to the people of
India for their generous and steady support for research in natural sciences, especially for string theory and cosmology.

%%%%%%%%%%%%%%%%%%%%%%%%%%%%%%%%%%%%%%%%%%%%%%%%%%%%%%%%%%%%%%%%%%%%%%%%%%%%%%%%%%%%%%%%%%%%%%%%%%%%%%%%%%%%%%%%%%%%%%%%%%%%%%%%%%%%%%%%%%%%%%%%%%%%%%%%%%%%%%%%%%%%%%%%%%%%%%%%%%%%%%%%%%%%%%%%%%%%%%%%%%%%%%%%%%%%%%%%%%%%%
%%%%%%%%%%%%%%%%%%%%%%%%%%%%%%%%%%%%%%%%%%%%%%%%%%%%%%%%%%%%%%%%%%%%%%%%%%%%%%%%%%%%%%%%%%%%%%%%%%%%%%%%%%%%%%%%%%%%%%%%%%%%%%%%%%%%%%%%%%%%%%%%%%%%%%%%%%%%%%%%%%%%%%%%%%%%%%%%%%%%%%%%%%%%%%%%%%%%%%%%

\end{document}